\newcommand\orc[1]{\href{https://orcid.org/#1}{\includegraphics[width=3mm]{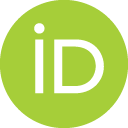}}}

\documentclass{aa}  

\usepackage{graphicx}
\usepackage{txfonts}
\usepackage[]{hyperref}
\usepackage{xcolor}  
\usepackage{bm}
\usepackage{graphicx}
\usepackage{txfonts}
\begin{document}

   \title{Inferring main-sequence stage and buoyancy-glitch amplitudes from Fourier spectra of gravity-mode period spacings}

   \subtitle{Ensemble Analysis of 26 Slowly Pulsating B Stars}
  
   \author{Zhao Guo\inst{1,2}\orc{0000-0002-0951-2171}
          \and
     Conny Aerts\inst{1,3,4}\orc{0000-0003-1822-7126}
     }
   \institute{Institute of Astronomy (IvS), Department of Physics and Astronomy, KU Leuven, Celestijnenlaan 200D, 3001 Leuven, Belgium\\
              \email{zhao.guo@kuleuven.be}
              \and
Department of Applied Mathematics, School of Mathematics, University of Leeds, Leeds LS2 9JT, UK
                          \and
             Department of Astrophysics, IMAPP, Radboud University Nijmegen,
PO Box 9010, 6500 GL Nijmegen, The Netherlands
\and
Max-Planck-Institut für Astronomie, Königstuhl 17, D-69117 
              Heidelberg, Germany
    \\          
    \email{conny.aerts@kuleuven.be}
}

   \date{Received Nov 15, 2025; accepted Dec 16, 2025}

 
  \abstract
   {Gravito-inertial-mode asteroseismology of intermediate-mass main-sequence stars took off with the 5-month uninterrupted light curves of the Convection, Rotation, and planeTary (CoRoT) space mission. It was developed in detail from the 4-year-long {\it Kepler\/} light curves, which provided a practical means to measure the rotation frequency in the transition layer between the convective core and the radiative envelope, where the local buoyancy frequency reaches a maximum. 
   Recently, a new buoyancy glitch inversion method based on the Fourier
spectra of gravity-mode period spacings was developed to probe that region further (Guo 2025). }
   {We aim to exploit the information contained in the variability of gravity-mode period spacings ($\Delta P$) in Slowly Pulsating B (SPB) stars with rotation. We investigate how well the main-sequence evolutionary stage can be inferred from this variability.}
   {We extract the frequency and amplitude of the variability in $\Delta P$ from the Fourier spectrum (FT). Both the period spacing $\Delta P$ and its periodic perturbations $\delta P$ (deviations from their asymptotic values) are used. The measured dominant frequency of $\Delta P$ allows us to infer the central hydrogen mass fraction, $X_c$, which is a main-sequence age indicator, using the approach in Guo (2025).}
   {The inferred $X_c$ values from $FT(\Delta P)$ mostly agree with previous results reported in the literature based on forward modelling of individual identified mode frequencies.
   We find that the buoyancy glitches $\delta N/N$ in SPB stars are generally less than $2\%$ in amplitude. Ensemble asteroseismic modeling of gravity-mode pulsators can now be carried out efficiently with our novel $FT(\Delta P)$ method once the internal rotation rate of the pulsators is known.} 
   {Our methodology offers a fast method for gravito-inertial asteroseismic applications in the era of ongoing and future large-scale, all-sky, space-based observations (to be) assembled with the Transiting Exoplanet Survey Satellite (TESS) and the PLAnetary Transits and Oscillations of stars (PLATO) space missions.}


\titlerunning{
The main-sequence stage and buoyancy-glitch amplitude from g-mode period spacings for 26 SPB stars}
\authorrunning{Z.\ Guo \& C.\ Aerts}

  \maketitle
%

\section{Introduction}

Slowly pulsating B (SPB) stars are mid- to late-B type main-sequence stars with masses between roughly $3 - 8 M_{\odot} $ that occupy a region above the classical instability strip on the Hertzsprung–Russell diagram \citep{Aer10}. They exhibit multi-periodic brightness variations with amplitudes of less than $\sim 50$ mmag and pulsation periods of about $0.5 - 5$ days, corresponding to high-radial-order, low-degree ($l=1,2$) gravity (g) modes which propagate in the radiative envelope
\citep{DeCatAerts2002}. SPB stars typically rotate at a moderate to high fraction of their critical velocity \citep{Pap17,Ped21}, making the effects of rotation important for their seismic interpretation \citep{Tow25}. Their pulsations are excited by the $\kappa$-mechanism operating in the iron-group opacity bump, which drives the g~modes \citep{Pamyatnykh1999,Tow25b,Szewczuk2017} and makes SPB stars key laboratories for studying stellar opacities \citep{Daszynska2010,Daszynska2020},
internal mixing \citep{Ped21,Ped22a}, and angular momentum transport in B-type stars \citep{Aer19,Ped22b,Mombarg2023,Aertsetal2025,Aerts2026}.


The asteroseismology of SPB stars heavily relies on pattern recognition of g~modes belonging to the same spherical degree $l$. 
It is based on the property that low-frequency g~modes of the same degree and azimuthal order, and of consecutive overtones, fulfill an asymptotic relation for the mode period spacings, $\Delta P$. These spacings are constant in a non-rotating, non-magnetic star but deviate from uniformity for real stars. Their period spacing values are proportional to 
the characteristic spacing $\Pi_0$ determined by the integral of the buoyancy frequency profile 
over the g-mode cavity
\citep[e.g.,][]{Miglio2008}. For a rotating pulsator, the values of $\Delta P$ are strongly dependent on the Coriolis force and thus allow for an estimate of the internal rotation rate at the position where the g~modes have dominant probing power
\citep[cf.,][for a review]{Bouabid2013,VanReeth2015b,AT2024}. 

Period spacing patterns of g~modes in SPB stars have first been found in the CoRoT data of two stars: the slowly rotating B3V star HD\,50230 \citep[][see Appendix\,A]{Deg10} and the rapidly rotating magnetic B3IV star HD\,43317 \citep{Pap12,Briquet2013,Buy18}. However, the real breakthrough in gravito-inertial asteroseismology came from the 4-year-long light curves assembled by the {\it Kepler\/} space telescope, which revealed   clear and long g-mode period spacings patterns in SPB stars, mostly of $l=1$ prograde modes \citep{Pap12, Pap14, Pap15, Pap17,Sze21,Ped21}. Similarly, nearly equally spaced pulsation period patterns have been found 
in numerous F-type pulsators called $\gamma\,$Dor stars \citep{VanReeth2015a,Li20}. While several {\it Kepler\/} $\gamma\,$Dor pulsators have been found in binary or multiple systems, this is the case for only a few SPB stars, e.g. HD\,201433 in \citet{Kal17},
KIC\,4931738 and KIC\,6352430 in 
\citet{Pap13}, and 
KIC\,4930889 in \citet{Pap17}.
However, several more are likely to be found in the near future from the large catalogue of TESS eclipsing binaries compiled by \citet{IJspeert2024a,IJspeert2024b}.

A step further than just using the measured value of $\Pi_0$ in stellar modelling 
is to perform individual mode frequency (or mode period) fitting \citep[see][for a methodological framework]{Aerts2018}. Such detailed seismic modeling of SPB pulsators has been 
pioneered by \citet{Mor15, Mor16} for the slow and moderate rotators KIC\,10526294 and KIC\,7760861, respectively. 
It has since also been performed 
for KIC\,3240411 by \citet{Sze18}, 
HD\,50230 by \citet{Wu19}, 
KIC\,8324482 by \citet{Wu20}, 
and the double-lined SPB binary KIC\,4930889 by \citet{Michielsen2023}. Each of these studies were conducted with a different level of detail and employed a variety of modelling tools and assumptions on the physics of stellar interiors.

Because applying methods such as those described in \citet{Aerts2018} to individual pulsators is time-consuming, \citet{Ped21} developed a statistical approach to
model the g-mode period spacings observed in 26 SPB stars. They found that stars exhibit diverse near-core mixing profiles -- both in strength and shape -- which significantly affect core-mass growth and stellar lifetimes, highlighting the need for proper calibrations of each of the individual active mixing processes.
They relied on statistical models to increase the resolution of theoretical period spacing values
as a function of stellar parameters such as mass
$M$, central hydrogen mass fraction $X_c$, convective core overshooting $f_{\rm ov}$, envelope mixing $D_{\rm env}$ and rotation rate $\Omega_{\rm rot}$. The results were further improved in \citet{Ped22a} with a focus on better assessment of uncertainties.

The aforementioned observed g-mode period spacings of SPB stars mostly exhibit variability induced by buoyancy glitches \citep{Cun24}. By focusing on slowly rotating SPB stars for which the Coriolis force can be ignored, \citet{Wu18} demonstrated that the frequencies with which period spacings vary ($f(\Delta P)$) scale approximately with the buoyancy-size ($\Lambda_{\mu})$ of the $\mu$-gradient region, making $f(\Delta P)$ a sensitive diagnostic of the 
central hydrogen fraction $X_c$ (hence main-sequence evolutionary stage). They proposed a “C-D like” diagram $f(\Delta P)$ vs mean period spacing $\langle \Delta P \rangle$ that, when combined with detailed modelling, allows constraints to be placed on both the stellar mass and evolutionary stage of SPB stars. 

\citet[hereafter G25]{Guo25} extended this work to include the Coriolis force
and derived an expression that links the Fourier spectrum of g-mode period spacings $FT(\Delta P)$ and of periodic perturbations $FT(\delta P)$ to buoyancy glitches $\delta N/N$. This approach is very general and can be used to infer any internal structure profiles with buoyancy glitches (e.g.\,from mass transfer as studied by \citet{Wu25} or mergers as in \citet{Henneco2025}'s theoretical predictions). It only applies to small glitches but does not assume any analytical description of the glitch shapes as in previous studies \citep{Cun19}. 
Using numerical MESA stellar models \citep{Pax11,Pax13} with varying mass, age, and convective-core overshooting, G25 found a tight, almost mass-independent linear relation between $X_c$ and $u_{\mu}$,
where this quantity represents the frequency of the variation in the period spacing similar to $f(\Delta P)$ in W18, but is dimensionless. He also showed that the amplitude of the variability in the period spacings is sensitive to the envelope mixing parameter $D_{\rm env}$. 
 



In this article, we perform an ensemble analysis of 
the g-mode period spacing variations for the 26 SPB stars derived by \citet{Ped21}. In Section 2, we discuss the overall properties of the Fourier spectra of g-mode period spacings. We then measure the frequencies and amplitudes of the variations (Sec. 3) and infer the central hydrogen mass fraction and buoyancy-glitch amplitudes using the theoretical predictions of G25 (Sec. 4). After discussing a few specific examples of SPBs in Sec. 5, we conclude in Sec. 6.

\section{Gravity-mode period spacings in SPB stars and their Fourier Spectra}

In the left panel of Figure 1, we show the gravity-mode period spacings of 26 SPBs in the observer's (inertial)
frame.
Most of the gravity modes presented (16 stars) are prograde dipole modes ($l=1, m=-1$ in the notation convention adopted in G25).
Nine stars have zonal dipole modes ($l=1, m=0$) and only one star (KIC 8714886) shows 
retrograde modes ($l=1, m=+1$).

To perform Fourier analysis, we apply the method of G25 and transform the periods 
and period spacings to the co-rotating frame by using the rotation rates reported in \citet{Ped21}, assuming uniform rotation (as listed in Table 1). The transformed period spacing series are displayed in the right panel of Figure 1.
The tilt of period spacings is essentially removed in the co-rotating frame, except for $m=0$ modes. Most of the period spacings show clear periodic variations.
In the last two columns of Table 1, we show the number of gravity modes in the series ($N$), which ranges from $N=6$ to $N=23$, and the rotation rates of the star $f_{\rm rot}$ in units of day$^{-1}$, respectively.

\begin{figure*}
   \centering    
 \includegraphics[width=17cm]{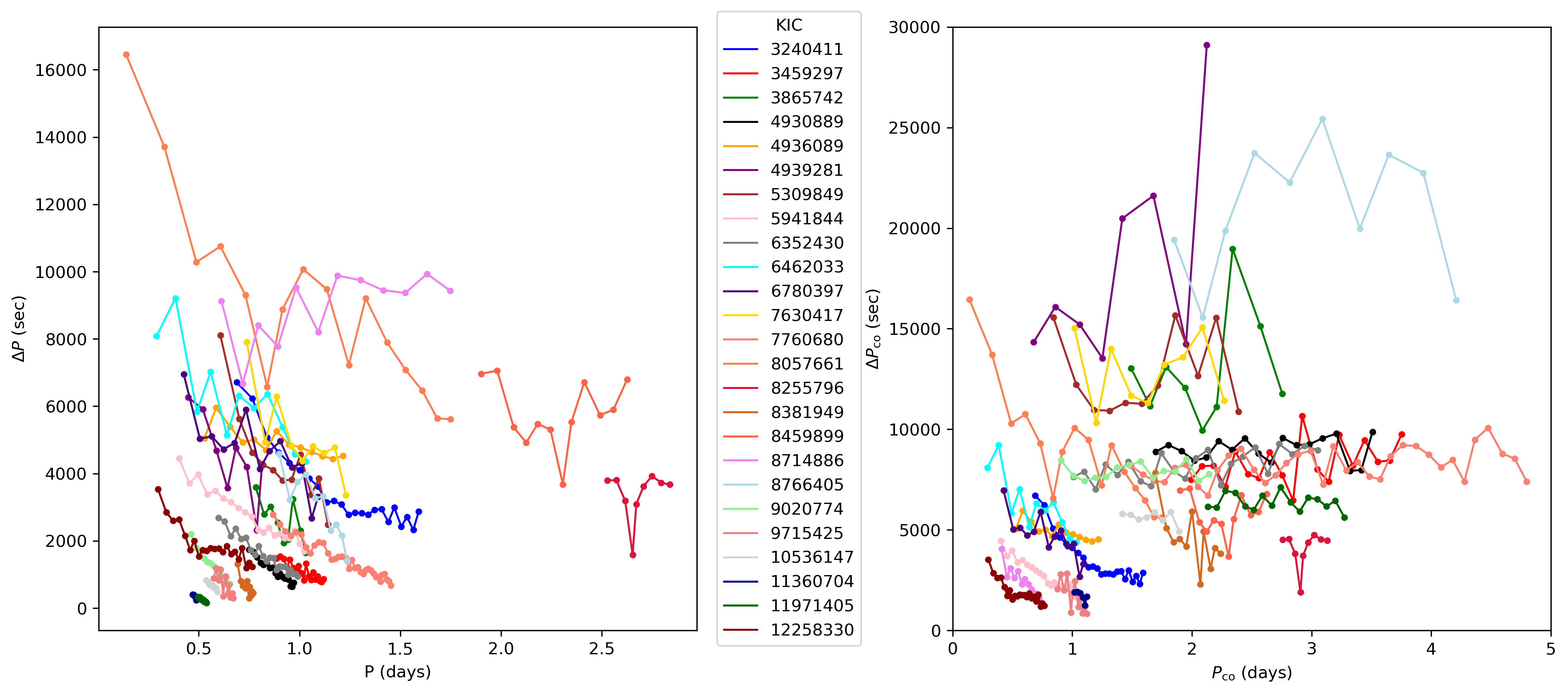}
   \caption{\textbf{Left:} Observed gravity-mode period spacings of 26 SPB stars in \citet{Ped21}.
   \textbf{Right:} Period spacings transformed to the co-rotating frame.}
   \label{FigDP}%
\end{figure*}

   \begin{figure*}
   \centering    
 \includegraphics[width=19cm]{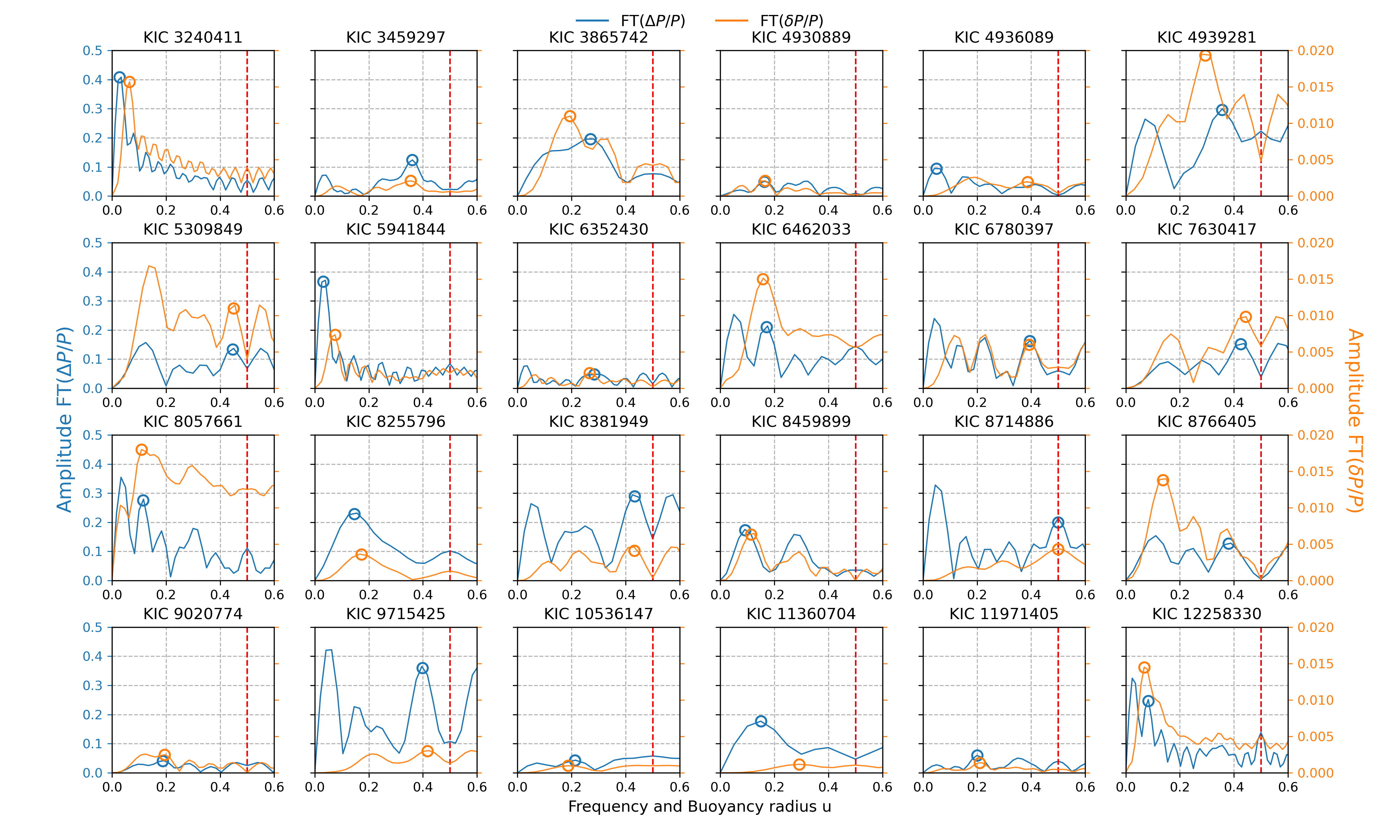}
   \caption{Fourier spectra of g-mode relative period spacings $\Delta P/P$ (blue)
   and period perturbations $\delta P/P$ (orange) for 24 SPB pulsators. The adopted peaks, which
   indicate the chemical transition points, are marked by circles.}
   \label{FigFT}%
    \end{figure*}

Figure 2 presents the Fourier spectra of the g-mode period spacings ($\Delta P$ normalized by its mean values and shown as blue curves) and of the relative period perturbations ($\delta P/P$, shown as orange curves) for 24 SPB stars in the sample. Similar Fourier spectra of the remaining two stars KIC 10526294 and KIC 7760680 were already reported in G25.
As in G25, we use the modes' radial orders (k) as `time' and period spacings/perturbations as `magnitude' of the variability.  Each panel corresponds to one star, with the horizontal axis showing the variation frequency (equivalent to the relative buoyancy radius $u$) and the vertical axis indicating the respective Fourier amplitudes. A vertical dashed red line at 0.5 marks the Nyquist frequency of the data. 
The $\Delta P/P$ spectra (blue) typically show clear peaks (indicated by blue circles) that trace the dominant periodicity in the g-mode spacing patterns. While most of the $\delta P/P$ spectra (orange) have peaks at similar locations (orange circles),
the low-frequency part often lacks the corresponding peaks seen in $\Delta P/P$.


Since we have far fewer data points compared to the regular Fourier spectra of {\it Kepler} light curves, the Fourier spectra exhibit only broad peaks. It is important to extract the correct peak that corresponds to the period-spacing variations.
We adopt the following criterion when selecting the peaks:
\textbf{(1)} We first select the highest and second highest peaks as our candidates.
\textbf{(2)} The peaks should appear in the Fourier spectra of both $\delta P/P$ and $\Delta P/P$. 
\textbf{(3)} We only consider the sub-Nyquist frequency region ($u<0.5$) of the spectrum. In principle, the dominant periodicity in period spacings can somewhat exceed 0.5 for very-late-stage SPB stars (see G25, Figure 3). In this sample, however, all the stars have their dominant variation frequencies at less than or equal to 0.5.

Since the Fourier amplitude spectrum $FT(\delta P/P)$ is equal to $|\delta N/N|$ to a good approximation,
the orange lines in Figure 2 also represent the absolute value of buoyancy glitch profiles.
Together, these results reveal diverse buoyancy 
signatures across the sample, ranging from stars with strong, narrow peaks - indicative of well-defined $\mu$ gradient regions - to others with multiple glitch features. Indeed, as shown in G25, Fourier spectra of MESA/GYRE \citep{Tow13,Sun23} pulsation periods also reveal the possibility of additional glitch locations (aside from the two dominant ones at the convective core boundary and at the $\mu$ gradient transition point). This diversity found from our method is fully in line with the shapes of the buoyancy frequency profiles deduced from forward modelling by \citet{Ped21}, as displayed in Fig.\,2 of \citet{Aerts2021-GIW}.

\section{Period spacing variations and their amplitudes inferred from the Fourier spectra}

\begin{figure}
   \centering    
 \includegraphics[width=\columnwidth]{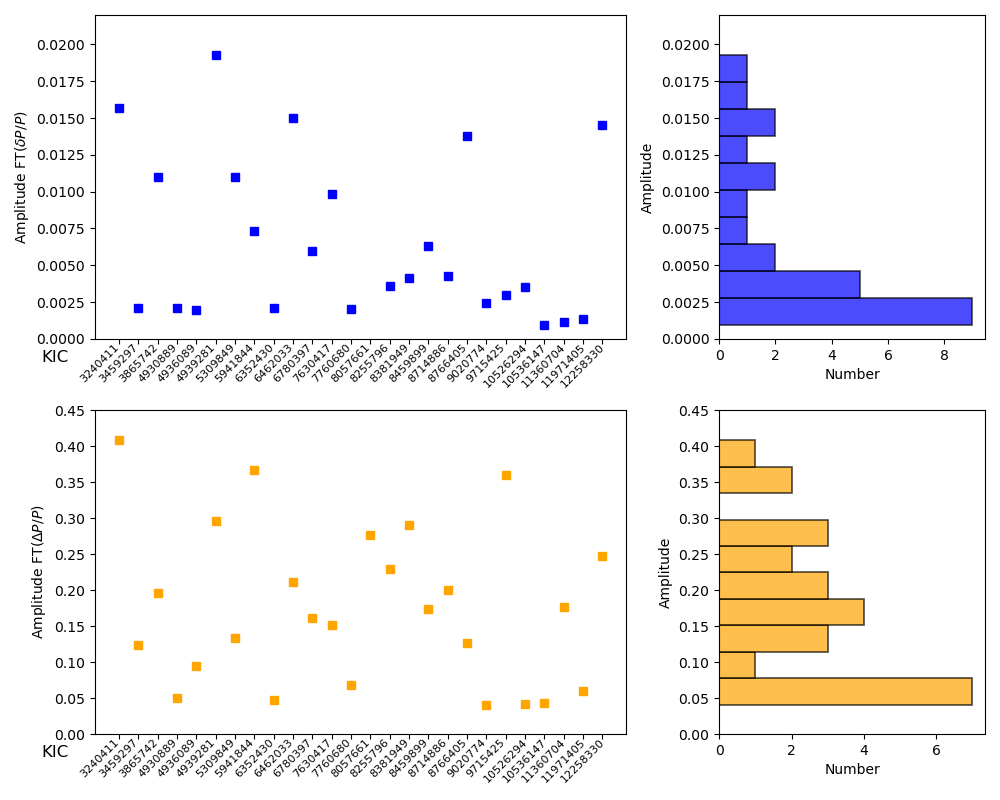}
   \caption{ \textbf{Lower: }Amplitudes of relative period spacings $FT(\Delta P/P)$ for 26 SPB stars. 
   \textbf{Upper:} Amplitudes of relative period perturbations $FT(\delta P/P)$, which are equal to 
   the amplitudes of buoyancy glitches $\delta N/N$. The right panels show the corresponding 
   histograms.}
      \label{Figamp}%
\end{figure}

Figure 3 shows the amplitude of $FT(\delta P/P)$ and $FT(\Delta P/P)$ for the 26 SPB stars in the sample on the left,
with the corresponding histograms in the right panel.
All the $FT(\delta P/P)$ amplitudes are lower than 2\%, with the bulk amplitude ranging from 0.002 to 0.007. This indicates that we are well within the perturbative (small glitch) regime. Since $FT(\delta P/P) \approx |\delta N/N|$, these measurements also correspond roughly with the actual relative buoyancy glitch amplitudes in the stars.

The amplitudes of $FT(\Delta P/P)$ range from about 0.05 to 0.40. The measured amplitudes ($A$) and variation frequencies ($u$) are tabulated in Table 1.
To calculate the uncertainties of variation frequencies and amplitudes, we use a Monte-Carlo bootstrap approach: we generate 200 synthetic realizations of the period-spacing time series by adding Gaussian noise to the measured values (consistent with their individual errors). For each realization, we compute the Fourier spectrum, measure the frequency, amplitude, and phase of the dominant peak, and then derive the $1 \sigma$ uncertainties from the resulting distributions of 200 measurements. In Table 1, we list these $1 \sigma$ uncertainties in parentheses, indicating the uncertainty on the final digit(s).

\section{Inferred central hydrogen mass fractions}

\begin{figure}
   \centering    
 \includegraphics[width=\columnwidth]{spb_Xc.png}
   \caption{Central hydrogen mass fraction $X_c$ of 26 SPB stars obtained by Pedersen et al. (2021) and (2022) (black and gray squares, respectively)
   and those inferred from Fourier spectra of period spacings (circles). Red symbols lie 
   within $2\sigma$ of the Pedersen et al. results, while purple symbols do not.}
         \label{FigXc}%
    \end{figure}

We can use the following calibrated relation between the variation frequency $u_{\mu}$ of period spacings and the central hydrogen mass fraction ($X_c$) from G25 to estimate the evolutionary stage of the 26 SPB stars:
\begin{equation}
   u_{\mu} = (X_c -0.70)/(-1.13).
\end{equation}
Our measured variation frequencies $u_{\mu}$ are reported in Table\,1 (both from $FT(\Delta P)$ and $FT(\delta P)$). In Figure 4, we compare the $X_c$ values derived from our Fourier analysis of period spacings (red/purple circles) with those obtained from the forward modelling by \citet{Ped21} (black squares) and \citet{Ped22a} (gray squares). Note that the results of \citet{Ped21} and \citet{Ped22a} are based on a quite sophisticated and time-consuming statistical seismic modeling approach, rather than detailed individual-frequency-fitting from seismic grid modeling. We have sorted the stars by their $X_c$ values in \citet{Ped21} for clarity.

Most of our $X_c$ values (20 stars, red symbols) are within $2 \sigma$ of the \citet{Ped21} results, whereas 6 stars (purple symbols) lie outside this range.
This encouraging result demonstrates the effectiveness of the $FT(\Delta P)$ method proposed in G25, provided that the internal rotation rotation frequency can be determined \citep[e.g.\,from the method by][]{VanReeth2016}. The simplicity of our method enables efficient large-scale ensemble seismic modeling of gravity-mode pulsators. 
Moreover, we find that for stars with detailed frequency-fitting seismic modeling, such as KIC 10526294 by \citet{Mor15} and KIC 7760680 by \citet{Mor16}, our results agree very well with those in these two papers. We adopt the more accurate results from these two modelling papers dedicated to these two stars in our Table 1 and figures instead of those in \citet{Ped21}.

\begin{table*}
\caption{Extracted frequencies ($u$) and amplitudes ($A$) of the period-spacing $\Delta P$ and period-perturbation $\delta P$ variability from the corresponding Fourier spectra shown in Figure 2. The other columns list the inferred central hydrogen fraction $X_c$, number of g modes used ($N$) and the rotation rates $f_{\rm rot}$ adopted from Pedersen et al., in units of cycles per day.}
\label{tab:peder_guoA}
\centering
\small
\setlength{\tabcolsep}{4pt}
\begin{tabular}{r r r r r r r r r}
\hline\hline
KIC & $u_{\Delta P}$ & $u_{\delta P}$ & $A_{\Delta P}$ & $A_{\delta P}$ & $X_{\rm c}$  & $N$  & $f_{\rm rot}$ (day$^{-1}$) \\
\hline
3240411 & 0.028(2) & 0.065(3) & 0.41(4) & 0.016(2) & 0.63(2) &  23 & 1.085\\
3459297 & 0.36(1) & 0.36(2) & 0.12(2) & 0.0021(3) & 0.29(5) &  19 & 0.627 \\
3865742 & 0.27(6) & 0.19(4) & 0.20(4) & 0.011(2) & 0.40(9) &  9 & 0.607\\
4930889 & 0.163(7) & 0.166(4) & 0.05(1) & 0.0021(3) & 0.35(4) &  18 & 0.745\\
4936089 & 0.050(3) & 0.39(3) & 0.09(2) & 0.0020(5) & 0.26(4) &  13 & 0.175\\
4939281 & 0.36(3) & 0.29(2) & 0.30(6) & 0.019(4) & 0.34(6) &  8 & 0.757\\
5309849 & 0.447(5) & 0.450(6) & 0.13(2) & 0.011(2) & 0.19(6) &  11 & 0.458\\
5941844 & 0.032(1) & 0.075(3) & 0.37(4) & 0.007(1) & 0.62(3) &  20 & 0.988\\
6352430 & 0.28(4) & 0.27(5) & 0.048(9) & 0.0021(4) & 0.38(3) &  22 & 0.681\\
6462033 & 0.171(6) & 0.16(2) & 0.21(4) & 0.015(3) & 0.51(4) &  11 & 0.656\\
6780397 & 0.395(9) & 0.392(9) & 0.16(5) & 0.006(1) & 0.25(4) &  13 & 0.527\\
7630417 & 0.425(8) & 0.44(1) & 0.15(3) & 0.010(2) & 0.20(7) &  9 & 0.372\\
7760680 & 0.19(2) & 0.19(2) & 0.068(9) & 0.0020(3) & 0.48(2) &  35 & 0.455 \\
8057661 & 0.12(2) & 0.11(3) & 0.28(5) & 0.05(1) & 0.57(4) &  16 & 0.489\\
8255796 & 0.15(3) & 0.17(2) & 0.23(4) & 0.0036(6) & 0.5(1) &  9 & 0.033\\
8381949 & 0.43(1) & 0.43(1) & 0.29(5) & 0.0041(7) & 0.21(7) &  11 & 0.852\\
8459899 & 0.092(7) & 0.114(6) & 0.17(3) & 0.006(1) & 0.60(5) & 12 & 0.103 \\
8714886 & 0.50(4) & 0.50(3) & 0.20(6) & 0.0043(9) & 0.14(4) &  12 & 0.814 \\
8766405 & 0.38(1) & 0.14(3) & 0.13(3) & 0.014(2) & 0.27(6) &  10 & 0.572\\
9020774 & 0.19(4) & 0.19(4) & 0.041(7) & 0.0025(5) & 0.49(3) &  14 &1.061\\
9715425 & 0.40(1) & 0.42(2) & 0.36(7) & 0.0030(5) & 0.23(4) &  13 & 0.598\\
10526294 & 0.059(6) & 0.078(5) & 0.042(7) & 0.0035(5) & 0.63(4) &  18 & 0.044\\
10536147 & 0.21(2) & 0.19(2) & 0.04(1) & 0.0010(2) & 0.49(7) &  8 & 1.159\\
11360704 & 0.15(5) & 0.29(4) & 0.18(4) & 0.0012(2) & 0.37(12) &  6 & 1.151\\
11971405 & 0.20(6) & 0.21(1) & 0.06(1) & 0.0014(3) & 0.47(4) &  16 & 1.551\\
12258330 & 0.08(1) & 0.07(2) & 0.25(5) & 0.014(2) & 0.61(2) & 22 &1.883\\
\hline
\end{tabular}
\tablefoot{%
$u_{\Delta P}, u_{\delta P}$ are the adopted frequencies of $\Delta P$ and $\delta P$ variability. 
$A_{\Delta P}, A_{\delta P}$ are the amplitudes of the variability. 
$X_{\rm c}$ denotes the central hydrogen mass fraction, and $\sigma X_{\rm c}$ its uncertainty. 
$N$ is the number of pulsation periods in the $\Delta P$ series. We report values with their $1 \sigma$ uncertainties in parentheses, indicating the uncertainty in the last quoted digit(s).
}
\end{table*}

\section{Discussion on selected SPB stars}

By examining the period spacing series and the fitting results in \citet{Ped21}, we identified several systems with poor agreement between our results and those in \citet{Ped21}. This may be due to several reasons, involving either the observational analyses or the forward modelling of the identified g~modes due to too uncertain input physics, or both. 

The observational analyses by \citet{Ped21} have  been revisited by both \citet{VanBeeck2021} and \citet{Patil2025}. Both these studies adopted different mathematical assumptions for the frequency analyses compared to \citet{Ped21}. While the latter study paid particular attention to the construction of period spacing patterns in one homogeneous way for the 26 SPB stars revisited here, this observable was not constructed for each of the stars in the two updated studies, where the focus was on the frequency analysis comparisons. 
This is why we rely on the results in \citet{Ped21} for our study.

\citet{VanBeeck2021} and \citet{Patil2025} in general found consistent results for the frequency values with \citet{Ped21} for the dominant modes in the patterns, but some pulsators have less certain identifications for the lower-amplitude modes in the patterns. This is particularly the case  for pulsators with a low variance reduction from harmonic fits to the {\it Kepler\/} light curve based on the significant frequencies \citep[see ][for an extensive discussion]{VanBeeck2021}. 

KIC\,8255796 represents a case where the observational frequency analyses results are consistent between the three studies by \citet{Ped21}, \citet{VanBeeck2021} and \citet{Patil2025}. However, \citet{Ped21} found only 8 data points in its pattern of identified modes and this observed pattern shows a large dip. The best forward models in \citet{Ped21} rather show small-amplitude variations and deviate significantly (up to $2000$ seconds) from the observed period spacings. This makes their derived $X_c$ value less reliable, as also illustrated by the largely different values in \citet{Ped21} and \citet{Ped22a}.
Similarly, the frequencies of the dominant modes in 
KIC\,8459899 are consistent between \citet{Ped21}, \citet{VanBeeck2021} and \citet{Patil2025} but 
its period spacing pattern shows one large, broad dip. None of the best forward models found by  \citet{Ped21} can reproduce this deep dip. For these two stars, the period-spacing variation is not strictly periodic, implying that our $FT(\Delta P)$ derived $X_c$ value should be considered as a  preliminary result in need of further scrutiny from more identified modes,

KIC\,3865742 is a very fast rotator. Our estimate of its evolutionary stage has a large uncertainty, making it deviate somewhat from the result by \citet{Ped21}. However the different result is not extreme, keeping in mind the error bar.
For KIC\,6462033 we also find a deviating value, but it is unsure what causes this. The best forward models by \citet{Ped21} seem of good quality and  
the frequencies of the dominant modes 
are consistent between \citet{Ped21}, \citet{VanBeeck2021} and \citet{Patil2025}. We do note that also this pulsator is an extremely fast rotator. The forward modelling of these two SPB stars by \citet{Ped21} relies on physics that ignores the oblateness of these stars due to their fast rotation and may explain the deviating values with our approach.

KIC\,8714886 has a somewhat deviating value for its evolutionary stage, but this is relatively modest keeping in mind the large difference in estimate between \citet{Ped21} and \citet{Ped22a}. This star is also a rather fast rotator and the only one with a pattern formed by retrograde modes, with a consistent structure between the results by \citet{Ped21} and \citet{Patil2025}. Its FT occurs exactly at the Nyquist frequency, which may make our estimate less reliable.

For KIC 11971405, even though the period spacings show only small-amplitude variations in \citet{Ped21}, \citet{Patil2025} hinted at a different pattern with a deep dip. This star is a fast rotator and the forward models in \citet{Ped21} 
all have vastly different $\Delta P$ values.
We thus conclude that our model-independent $FT(\Delta P)$ derived $X_c$ is likely more robust.

One star, KIC 8057661, shows $\delta P/P \approx 0.05$, which is an exceptional outlier. All other stars in the sample have $\delta P/P < 0.02$. 
This SPB has a very complex amplitude spectrum as revealed in all three studies by 
\citet{Ped21}, \citet{VanBeeck2021}, and \citet{Patil2025}.
This may be due to a data-reduction issue, so we consider this measurement unreliable. Its Fourier spectrum $FT(\delta P/P)$ shown in Figure 2 is scaled by a factor of $1/3$, and its amplitude value is not shown in Figure 3.

Observations of SPB pulsators reveal signatures of magnetic fields in a few class members \citep{Buy18, Buy19}. A strong magnetic field can even suppress the gravity modes \citep{Lec22}. In particular, \citet{Buy18} performed detailed modelling of the g-mode period spacings in the CoRoT target HD\,43317. In terms of period spacing variations, the magnetic field bends the period-spacing pattern slightly \citep{Dhouib2022,Rui2024}, while it does not affect the higher-frequency variability, thereby inducing a very low-frequency peak in the $FT(\Delta P)$. Thus the inferred $X_c$ may be slightly biased for very young SPB stars. In general, we expect that magnetic effects can be neglected in our analysis of this SPB-star sample compared to other effects. Indeed, 
we note that we used Eq.\,(1) with an initial hydrogen mass fraction of $X_{init}=0.71$ to infer the $X_c$ value. This expression depends on the chemical composition 
of the MESA models used by G25, who relied on the solar values in
\citet{GrevesseNoels1998}.
The eight model grids computed by \citet{Ped21} rather relied on the \citet{Asplund2009} chemical composition for the Sun.
This may affect the estimate of the evolutionary state based on the inferred hydrogen mass fraction somewhat. We also point out that the envelope mixing coefficient $D_{\rm env}$ has an effect on the deduced $X_{\rm c}$ values for this SPB sample. Taking these two differences between G25 and \citet{Ped21} into account, we find overall good results between our method and the literature values for the evolutionary stage of most of the SPB stars.

\section{Conclusions}

By using a sample of 26 SPB stars, we verify the validity of the $FT(\Delta P)$ approach in G25. Using the tight relation between the variability of gravity-mode period spacings 
      and the main-sequence stage (that is, the central hydrogen mass fraction $X_c$),
      our inferred $X_c$ values mostly (for about $70\%$) agree with previous statistical seismic modeling by \citet{Ped21}.
      
The Fourier spectra of g-mode period spacings and their periodic perturbations indicate
      that SPB stars have buoyancy glitches in their Brunt--V\"ais\"al\"a\ profile with relative amplitudes $\delta N/N$
       less than $\approx 2\%$. The bulk glitch amplitude is about $0.2-1 \%$.

The $FT(\Delta P)$ method provides a rapid and efficient means for performing ensemble seismic modeling of gravity-mode pulsators, making it especially valuable in the era of space-based data.
It requires the internal rotation frequency in the zone of the Brunt--V\"ais\"al\"a\ peak just outside the convective core 
to be known with good precision, in order to transform the observed period-spacing patterns to the co-rotating frame. 

Future work can refine the method we developed and applied here, in order to take into account a variety of chemical compositions and levels of envelope mixing, which were not considered in G25. 


\begin{acknowledgements}
      The authors acknowledge support from the European Research Council (ERC) under the Horizon Europe program (Synergy Grant agreement 101071505: 4D-STAR). While partially funded by the European Union, views and opinions expressed are however those of the author only and do not necessarily reflect those of the European Union or the European Research Council. Neither the European Union nor the granting authority can be held responsible for them. ZG is also supported by STFC grant UKRI1179.
\end{acknowledgements}

\begin{appendix}
\section{Fourier analysis of the first SPB star with detected period-spacing variations}

\begin{figure}
\centering    
\includegraphics[width=\columnwidth]{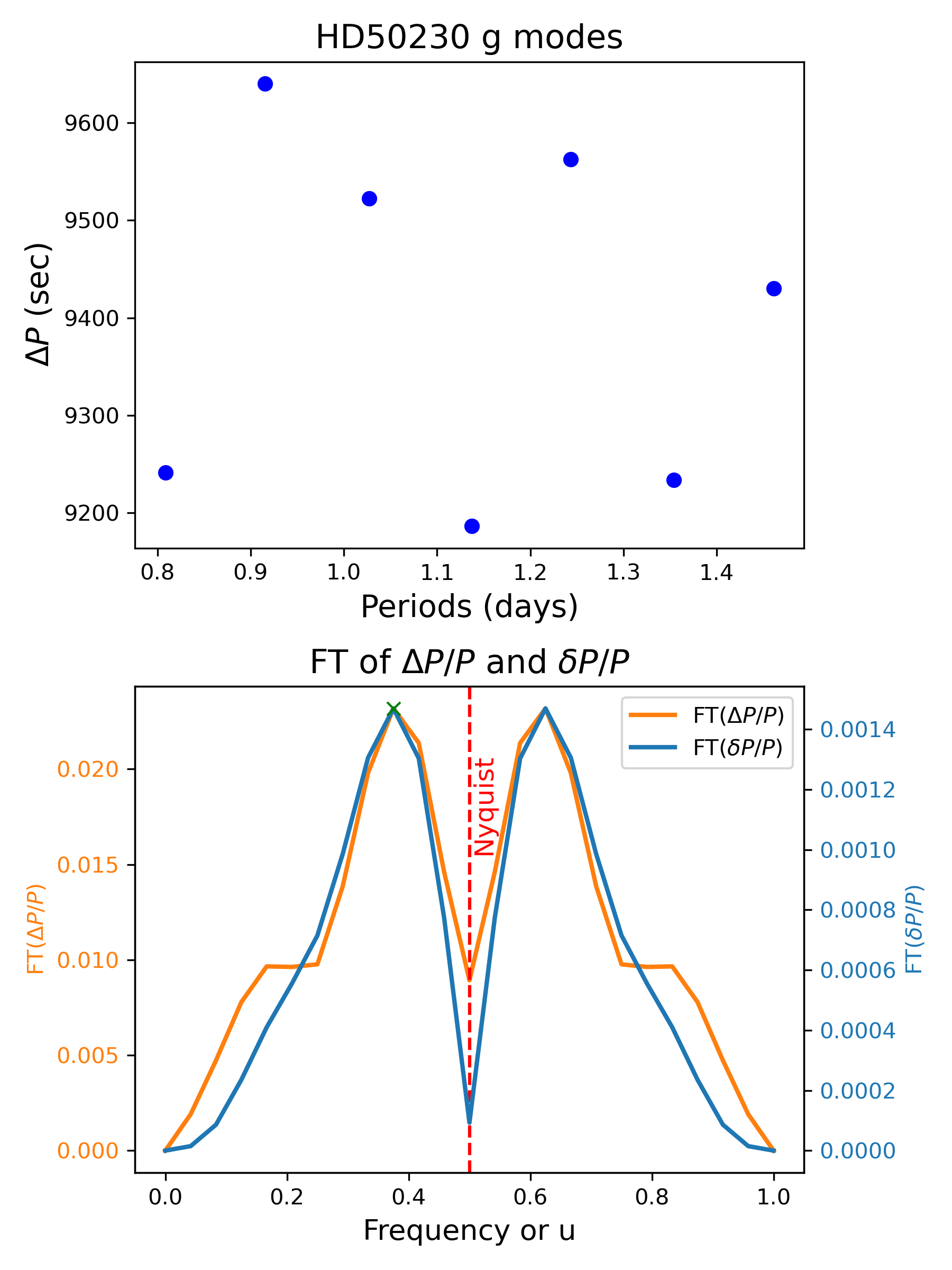}
\caption{\textit{Upper}: Observed g-mode period spacings in HD 50230. \textit{Lower}: Fourier spectra of relative period spacings $\Delta P/P$ and relative pulsation-period perturbations $\delta P/P$. The dominant peak is at $u=0.375$, indicating $X_c \approx 0.28$.} 
\label{hdft}%

\end{figure}

We show the g-mode period spacing analysis of the first SPB star with observed $\Delta P$ variations. This concerns the slowly rotating SPB star HD\,50230 observed by the CoRoT satellite during 5 months.
The upper panel of Fig.\,\ref{hdft} 
shows the observed period spacing - period diagram as in \citet{Deg10}.
The lower panel presents the Fourier spectra of $\Delta P/P$ and $\delta P/P$ in orange and blue, respectively. Both spectra exhibit a similar feature, with the dominant periodicity located $u=0.375$ (labeled by a green cross). This indicates a late-main-sequence hydrogen mass fraction $X_c \approx 0.28$ (from Eq. 1). This is in very good agreement with \citet{Wu19}, who found $X_{\rm c}=0.298$.
The $FT(\delta P/P)$ has a peak amplitude of $0.00147$, indicating a $\delta N/N$ of the same amplitude ($\sim 0.15\%$).  

\end{appendix}

\bibliographystyle{aa} 
\bibliography{spb} 

\begin{thebibliography}{60}
\expandafter\ifx\csname natexlab\endcsname\relax\def\natexlab#1{#1}\fi

\bibitem[{{Aerts}(2026)}]{Aerts2026}
{Aerts}, C. 2026, \aap, in press, arXiv:2511.02909

\bibitem[{{Aerts} {et~al.}(2021){Aerts}, {Augustson}, {Mathis}, {Pedersen}, {Mombarg}, {Vanlaer}, {Van Beeck}, \& {Van Reeth}}]{Aerts2021-GIW}
{Aerts}, C., {Augustson}, K., {Mathis}, S., {et~al.} 2021, \aap, 656, A121

\bibitem[{{Aerts} {et~al.}(2010){Aerts}, {Christensen-Dalsgaard}, \& {Kurtz}}]{Aer10}
{Aerts}, C., {Christensen-Dalsgaard}, J., \& {Kurtz}, D.~W. 2010, {Asteroseismology, Springer-Verlag, Heidelberg, Germany}

\bibitem[{{Aerts} {et~al.}(2019){Aerts}, {Mathis}, \& {Rogers}}]{Aer19}
{Aerts}, C., {Mathis}, S., \& {Rogers}, T.~M. 2019, \araa, 57, 35

\bibitem[{{Aerts} {et~al.}(2018){Aerts}, {Molenberghs}, {Michielsen}, {Pedersen}, {Bj{\"o}rklund}, {Johnston}, {Mombarg}, {Bowman}, {Buysschaert}, {P{\'a}pics}, {Sekaran}, {Sundqvist}, {Tkachenko}, {Truyaert}, {Van Reeth}, \& {Vermeyen}}]{Aerts2018}
{Aerts}, C., {Molenberghs}, G., {Michielsen}, M., {et~al.} 2018, \apjs, 237, 15

\bibitem[{{Aerts} \& {Tkachenko}(2024)}]{AT2024}
{Aerts}, C. \& {Tkachenko}, A. 2024, \aap, 692, R1

\bibitem[{{Aerts} {et~al.}(2025){Aerts}, {Van Reeth}, {Mombarg}, \& {Hey}}]{Aertsetal2025}
{Aerts}, C., {Van Reeth}, T., {Mombarg}, J. S.~G., \& {Hey}, D. 2025, \aap, 695, A214

\bibitem[{{Asplund} {et~al.}(2009){Asplund}, {Grevesse}, {Sauval}, \& {Scott}}]{Asplund2009}
{Asplund}, M., {Grevesse}, N., {Sauval}, A.~J., \& {Scott}, P. 2009, \araa, 47, 481

\bibitem[{{Bouabid} {et~al.}(2013){Bouabid}, {Dupret}, {Salmon}, {Montalb{\'a}n}, {Miglio}, \& {Noels}}]{Bouabid2013}
{Bouabid}, M.-P., {Dupret}, M.-A., {Salmon}, S., {et~al.} 2013, \mnras, 429, 2500

\bibitem[{{Briquet} {et~al.}(2013){Briquet}, {Neiner}, {Leroy}, {P{\'a}pics}, \& {MiMeS Collaboration}}]{Briquet2013}
{Briquet}, M., {Neiner}, C., {Leroy}, B., {P{\'a}pics}, P.~I., \& {MiMeS Collaboration}. 2013, \aap, 557, L16

\bibitem[{{Buysschaert} {et~al.}(2018){Buysschaert}, {Aerts}, {Bowman}, {Johnston}, {Van Reeth}, {Pedersen}, {Mathis}, \& {Neiner}}]{Buy18}
{Buysschaert}, B., {Aerts}, C., {Bowman}, D.~M., {et~al.} 2018, \aap, 616, A148

\bibitem[{{Buysschaert} {et~al.}(2019){Buysschaert}, {Neiner}, {Martin}, {Oksala}, {Aerts}, {Tkachenko}, {Alecian}, \& {MiMeS Collaboration}}]{Buy19}
{Buysschaert}, B., {Neiner}, C., {Martin}, A.~J., {et~al.} 2019, \aap, 622, A67

\bibitem[{{Cunha} {et~al.}(2019){Cunha}, {Avelino}, {Christensen-Dalsgaard}, {Stello}, {Vrard}, {Jiang}, \& {Mosser}}]{Cun19}
{Cunha}, M.~S., {Avelino}, P.~P., {Christensen-Dalsgaard}, J., {et~al.} 2019, \mnras, 490, 909

\bibitem[{{Cunha} {et~al.}(2024){Cunha}, {Damasceno}, {Amaral}, {Falorca}, {Christensen-Dalsgaard}, \& {Avelino}}]{Cun24}
{Cunha}, M.~S., {Damasceno}, Y.~C., {Amaral}, J., {et~al.} 2024, \aap, 687, A100

\bibitem[{{Daszy{\'n}ska-Daszkiewicz} \& {Walczak}(2010)}]{Daszynska2010}
{Daszy{\'n}ska-Daszkiewicz}, J. \& {Walczak}, P. 2010, \mnras, 403, 496

\bibitem[{{Daszy{\'n}ska-Daszkiewicz} {et~al.}(2020){Daszy{\'n}ska-Daszkiewicz}, {Walczak}, {Pamyatnykh}, \& {Szewczuk}}]{Daszynska2020}
{Daszy{\'n}ska-Daszkiewicz}, J., {Walczak}, P., {Pamyatnykh}, A., \& {Szewczuk}, W. 2020, in XXXIX Polish Astronomical Society Meeting, ed. K.~{Ma{\l}ek}, M.~{Poli{\'n}ska}, A.~{Majczyna}, G.~{Stachowski}, R.~{Poleski}, {\L}.~{Wyrzykowski}, \& A.~{R{\'o}{\.z}a{\'n}ska}, Vol.~10, 136--141

\bibitem[{{De Cat} \& {Aerts}(2002)}]{DeCatAerts2002}
{De Cat}, P. \& {Aerts}, C. 2002, \aap, 393, 965

\bibitem[{{Degroote} {et~al.}(2010){Degroote}, {Aerts}, {Baglin}, {Miglio}, {Briquet}, {Noels}, {Niemczura}, {Montalban}, {Bloemen}, {Oreiro}, {Vu{\v{c}}kovi{\'c}}, {Smolders}, {Auvergne}, {Baudin}, {Catala}, \& {Michel}}]{Deg10}
{Degroote}, P., {Aerts}, C., {Baglin}, A., {et~al.} 2010, \nat, 464, 259

\bibitem[{{Dhouib} {et~al.}(2022){Dhouib}, {Mathis}, {Bugnet}, {Van Reeth}, \& {Aerts}}]{Dhouib2022}
{Dhouib}, H., {Mathis}, S., {Bugnet}, L., {Van Reeth}, T., \& {Aerts}, C. 2022, \aap, 661, A133

\bibitem[{{Grevesse} \& {Sauval}(1998)}]{GrevesseNoels1998}
{Grevesse}, N. \& {Sauval}, A.~J. 1998, \ssr, 85, 161

\bibitem[{{Guo}(2025)}]{Guo25}
{Guo}, Z. 2025, arXiv e-prints, arXiv:2511.05780

\bibitem[{{Henneco} {et~al.}(2025){Henneco}, {Schneider}, {Heller}, {Hekker}, \& {Aerts}}]{Henneco2025}
{Henneco}, J., {Schneider}, F.~R.~N., {Heller}, M., {Hekker}, S., \& {Aerts}, C. 2025, \aap, 698, A49

\bibitem[{{IJspeert} {et~al.}(2024{\natexlab{a}}){IJspeert}, {Tkachenko}, {Johnston}, \& {Aerts}}]{IJspeert2024a}
{IJspeert}, L.~W., {Tkachenko}, A., {Johnston}, C., \& {Aerts}, C. 2024{\natexlab{a}}, \aap, 691, A242

\bibitem[{{IJspeert} {et~al.}(2024{\natexlab{b}}){IJspeert}, {Tkachenko}, {Johnston}, {Pr{\v{s}}a}, {Wells}, \& {Aerts}}]{IJspeert2024b}
{IJspeert}, L.~W., {Tkachenko}, A., {Johnston}, C., {et~al.} 2024{\natexlab{b}}, \aap, 685, A62

\bibitem[{{Kallinger} {et~al.}(2017){Kallinger}, {Weiss}, {Beck}, {Pigulski}, {Kuschnig}, {Tkachenko}, {Pakhomov}, {Ryabchikova}, {L{\"u}ftinger}, {Palle}, {Semenko}, {Handler}, {Koudelka}, {Matthews}, {Moffat}, {Pablo}, {Popowicz}, {Rucinski}, {Wade}, \& {Zwintz}}]{Kal17}
{Kallinger}, T., {Weiss}, W.~W., {Beck}, P.~G., {et~al.} 2017, \aap, 603, A13

\bibitem[{{Lecoanet} {et~al.}(2022){Lecoanet}, {Bowman}, \& {Van Reeth}}]{Lec22}
{Lecoanet}, D., {Bowman}, D.~M., \& {Van Reeth}, T. 2022, \mnras, 512, L16

\bibitem[{{Li} {et~al.}(2020){Li}, {Van Reeth}, {Bedding}, {Murphy}, {Antoci}, {Ouazzani}, \& {Barbara}}]{Li20}
{Li}, G., {Van Reeth}, T., {Bedding}, T.~R., {et~al.} 2020, \mnras, 491, 3586

\bibitem[{{Michielsen} {et~al.}(2023){Michielsen}, {Van Reeth}, {Tkachenko}, \& {Aerts}}]{Michielsen2023}
{Michielsen}, M., {Van Reeth}, T., {Tkachenko}, A., \& {Aerts}, C. 2023, \aap, 679, A6

\bibitem[{{Miglio} {et~al.}(2008){Miglio}, {Montalb{\'a}n}, {Noels}, \& {Eggenberger}}]{Miglio2008}
{Miglio}, A., {Montalb{\'a}n}, J., {Noels}, A., \& {Eggenberger}, P. 2008, \mnras, 386, 1487

\bibitem[{{Mombarg}(2023)}]{Mombarg2023}
{Mombarg}, J.~S.~G. 2023, \aap, 677, A63

\bibitem[{{Moravveji} {et~al.}(2015){Moravveji}, {Aerts}, {P{\'a}pics}, {Triana}, \& {Vandoren}}]{Mor15}
{Moravveji}, E., {Aerts}, C., {P{\'a}pics}, P.~I., {Triana}, S.~A., \& {Vandoren}, B. 2015, \aap, 580, A27

\bibitem[{{Moravveji} {et~al.}(2016){Moravveji}, {Townsend}, {Aerts}, \& {Mathis}}]{Mor16}
{Moravveji}, E., {Townsend}, R. H.~D., {Aerts}, C., \& {Mathis}, S. 2016, \apj, 823, 130

\bibitem[{{Pamyatnykh}(1999)}]{Pamyatnykh1999}
{Pamyatnykh}, A.~A. 1999, \actaa, 49, 119

\bibitem[{{P{\'a}pics} {et~al.}(2012){P{\'a}pics}, {Briquet}, {Baglin}, {Poretti}, {Aerts}, {Degroote}, {Tkachenko}, {Morel}, {Zima}, {Niemczura}, {Rainer}, {Hareter}, {Baudin}, {Catala}, {Michel}, {Samadi}, \& {Auvergne}}]{Pap12}
{P{\'a}pics}, P.~I., {Briquet}, M., {Baglin}, A., {et~al.} 2012, \aap, 542, A55

\bibitem[{{P{\'a}pics} {et~al.}(2014){P{\'a}pics}, {Moravveji}, {Aerts}, {Tkachenko}, {Triana}, {Bloemen}, \& {Southworth}}]{Pap14}
{P{\'a}pics}, P.~I., {Moravveji}, E., {Aerts}, C., {et~al.} 2014, \aap, 570, A8

\bibitem[{{P{\'a}pics} {et~al.}(2013){P{\'a}pics}, {Tkachenko}, {Aerts}, {Briquet}, {Marcos-Arenal}, {Beck}, {Uytterhoeven}, {Trivi{\~n}o Hage}, {Southworth}, {Clubb}, {Bloemen}, {Degroote}, {Jackiewicz}, {McKeever}, {Van Winckel}, {Niemczura}, {Gameiro}, \& {Debosscher}}]{Pap13}
{P{\'a}pics}, P.~I., {Tkachenko}, A., {Aerts}, C., {et~al.} 2013, \aap, 553, A127

\bibitem[{{P{\'a}pics} {et~al.}(2015){P{\'a}pics}, {Tkachenko}, {Aerts}, {Van Reeth}, {De Smedt}, {Hillen}, {{\O}stensen}, \& {Moravveji}}]{Pap15}
{P{\'a}pics}, P.~I., {Tkachenko}, A., {Aerts}, C., {et~al.} 2015, \apjl, 803, L25

\bibitem[{{P{\'a}pics} {et~al.}(2017){P{\'a}pics}, {Tkachenko}, {Van Reeth}, {Aerts}, {Moravveji}, {Van de Sande}, {De Smedt}, {Bloemen}, {Southworth}, {Debosscher}, {Niemczura}, \& {Gameiro}}]{Pap17}
{P{\'a}pics}, P.~I., {Tkachenko}, A., {Van Reeth}, T., {et~al.} 2017, \aap, 598, A74

\bibitem[{{Patil} {et~al.}(2025){Patil}, {Aerts}, {Wang}, {Van Beeck}, \& {Pedersen}}]{Patil2025}
{Patil}, A., {Aerts}, C., {Wang}, N., {Van Beeck}, J., \& {Pedersen}, M.~G. 2025, \aap, submitted

\bibitem[{{Paxton} {et~al.}(2011){Paxton}, {Bildsten}, {Dotter}, {Herwig}, {Lesaffre}, \& {Timmes}}]{Pax11}
{Paxton}, B., {Bildsten}, L., {Dotter}, A., {et~al.} 2011, \apjs, 192, 3

\bibitem[{{Paxton} {et~al.}(2013){Paxton}, {Cantiello}, {Arras}, {Bildsten}, {Brown}, {Dotter}, {Mankovich}, {Montgomery}, {Stello}, {Timmes}, \& {Townsend}}]{Pax13}
{Paxton}, B., {Cantiello}, M., {Arras}, P., {et~al.} 2013, \apjs, 208, 4

\bibitem[{{Pedersen}(2022{\natexlab{a}})}]{Ped22b}
{Pedersen}, M.~G. 2022{\natexlab{a}}, \apj, 940, 49

\bibitem[{{Pedersen}(2022{\natexlab{b}})}]{Ped22a}
{Pedersen}, M.~G. 2022{\natexlab{b}}, \apj, 930, 94

\bibitem[{{Pedersen} {et~al.}(2021){Pedersen}, {Aerts}, {P{\'a}pics}, {Michielsen}, {Gebruers}, {Rogers}, {Molenberghs}, {Burssens}, {Garcia}, \& {Bowman}}]{Ped21}
{Pedersen}, M.~G., {Aerts}, C., {P{\'a}pics}, P.~I., {et~al.} 2021, Nature Astronomy, 5, 715

\bibitem[{{Rui} {et~al.}(2024){Rui}, {Ong}, \& {Mathis}}]{Rui2024}
{Rui}, N.~Z., {Ong}, J.~M.~J., \& {Mathis}, S. 2024, \mnras, 527, 6346

\bibitem[{{Sun} {et~al.}(2023){Sun}, {Townsend}, \& {Guo}}]{Sun23}
{Sun}, M., {Townsend}, R.~H.~D., \& {Guo}, Z. 2023, \apj, 945, 43

\bibitem[{{Szewczuk} \& {Daszy{\'n}ska-Daszkiewicz}(2017)}]{Szewczuk2017}
{Szewczuk}, W. \& {Daszy{\'n}ska-Daszkiewicz}, J. 2017, \mnras, 469, 13

\bibitem[{{Szewczuk} \& {Daszy{\'n}ska-Daszkiewicz}(2018)}]{Sze18}
{Szewczuk}, W. \& {Daszy{\'n}ska-Daszkiewicz}, J. 2018, \mnras, 478, 2243

\bibitem[{{Szewczuk} {et~al.}(2021){Szewczuk}, {Walczak}, \& {Daszy{\'n}ska-Daszkiewicz}}]{Sze21}
{Szewczuk}, W., {Walczak}, P., \& {Daszy{\'n}ska-Daszkiewicz}, J. 2021, \mnras, 503, 5894

\bibitem[{{Townsend}(2005{\natexlab{a}})}]{Tow25}
{Townsend}, R.~H.~D. 2005{\natexlab{a}}, \mnras, 360, 465

\bibitem[{{Townsend}(2005{\natexlab{b}})}]{Tow25b}
{Townsend}, R.~H.~D. 2005{\natexlab{b}}, \mnras, 364, 573

\bibitem[{{Townsend} \& {Teitler}(2013)}]{Tow13}
{Townsend}, R.~H.~D. \& {Teitler}, S.~A. 2013, \mnras, 435, 3406

\bibitem[{{Van Beeck} {et~al.}(2021){Van Beeck}, {Bowman}, {Pedersen}, {Van Reeth}, {Van Hoolst}, \& {Aerts}}]{VanBeeck2021}
{Van Beeck}, J., {Bowman}, D.~M., {Pedersen}, M.~G., {et~al.} 2021, \aap, 655, A59

\bibitem[{{Van Reeth} {et~al.}(2016){Van Reeth}, {Tkachenko}, \& {Aerts}}]{VanReeth2016}
{Van Reeth}, T., {Tkachenko}, A., \& {Aerts}, C. 2016, \aap, 593, A120

\bibitem[{{Van Reeth} {et~al.}(2015{\natexlab{a}}){Van Reeth}, {Tkachenko}, {Aerts}, {P{\'a}pics}, {Degroote}, {Debosscher}, {Zwintz}, {Bloemen}, {De Smedt}, {Hrudkova}, {Raskin}, \& {Van Winckel}}]{VanReeth2015b}
{Van Reeth}, T., {Tkachenko}, A., {Aerts}, C., {et~al.} 2015{\natexlab{a}}, \aap, 574, A17

\bibitem[{{Van Reeth} {et~al.}(2015{\natexlab{b}}){Van Reeth}, {Tkachenko}, {Aerts}, {P{\'a}pics}, {Triana}, {Zwintz}, {Degroote}, {Debosscher}, {Bloemen}, {Schmid}, {De Smedt}, {Fremat}, {Fuentes}, {Homan}, {Hrudkova}, {Karjalainen}, {Lombaert}, {Nemeth}, {{\O}stensen}, {Van De Steene}, {Vos}, {Raskin}, \& {Van Winckel}}]{VanReeth2015a}
{Van Reeth}, T., {Tkachenko}, A., {Aerts}, C., {et~al.} 2015{\natexlab{b}}, \apjs, 218, 27

\bibitem[{{Wu} {et~al.}(2025){Wu}, {Guo}, \& {Li}}]{Wu25}
{Wu}, T., {Guo}, Z., \& {Li}, Y. 2025, arXiv e-prints, arXiv:2511.14372

\bibitem[{{Wu} \& {Li}(2019)}]{Wu19}
{Wu}, T. \& {Li}, Y. 2019, \apj, 881, 86

\bibitem[{{Wu} {et~al.}(2018){Wu}, {Li}, \& {Deng}}]{Wu18}
{Wu}, T., {Li}, Y., \& {Deng}, Z.-m. 2018, \apj, 867, 47

\bibitem[{{Wu} {et~al.}(2020){Wu}, {Li}, {Deng}, {Lin}, {Song}, \& {Jiang}}]{Wu20}
{Wu}, T., {Li}, Y., {Deng}, Z.-m., {et~al.} 2020, \apj, 899, 38

\end{thebibliography}

\end{document}